# All-optical transistor action by off-resonant activation at laser threshold

**David L. Andrews[1],* and David S. Bradshaw[1]**

[1]*Nanostructures and Photomolecular Systems, School of Chemistry,*

*University of East Anglia, Norwich NR4 7TJ, United Kingdom*

*Corresponding author: david.andrews@physics.org*

The development of viable all-optical data processing systems has immense importance for both the computing and telecommunication industries, but device realization remains elusive. In this Letter, we propose an innovative mechanism deployed as a basis for all-optical transistor action. In detail, it is determined that an optically pumped system, operating just below laser threshold, can exhibit a greatly enhanced output on application of an off-resonant beam of sufficient intensity. The electrodynamics of the underlying, nonlinear optical mechanism is analyzed, model calculations are performed, and the results are illustrated graphically.

*OCIS codes: (020.5580) Quantum electrodynamics; (140.3460) Lasers;*

In the pursuit of improved platforms for computing, communications and internet connectivity, the numerous potential advantages of all-optical systems are well-known. Key amongst the anticipated benefits, as compared to current electronic implementations, are a greatly increased speed and fidelity of data transmission, and reduced energy losses – the latter feature having additional significance for the global energy-management agenda. With a diverse range of sources and fibre optical connections already in production, much current effort is being devoted towards forging optical components such as an all-optical transistor. The underlying principle of any such device, like its electronic antecedent, is to effect the switching or amplification of a source, under the control of a signal input. One novel scheme for the realization of such a device, based on saturated absorption, has very recently been proposed by Hwang *et al.* [1]. Other, all-optical switching systems that have recently been entertained are based on electromagnetic induced transparency (EIT) [2-5], the optical Kerr effect [6-8], nonlinear transmission through coupling with surface plasmons [9-11], and beam filament rotation by application of a signal beam [12,13].

It can be anticipated that the desired characteristics of any practicable all-optical transistor will depend crucially on the engagement of a strongly nonlinear optical response. The scheme that is proposed in the present work is based upon a third-order nonlinearity – its effect enhanced by stimulated emission – operating within a system designed to exploit the highly nonlinear response observed in a system at the threshold for laser emission. In contrast to the work of Hwang *et al.*, stimulated emission here primarily occurs in the course of forward scattering by a throughput beam whose optical frequency is purposely off-set from resonance. Detailed analysis shows that this beam, acting as the input signal, can modify the kinetics of emission and so lead to a dramatically enhanced output [14]. The following results of calculations, for three-level laser systems, highlight the significant potential for device development.

Consider a typical three-level laser system optically pumped within a microcavity. The kinetics of emission are primarily determined by a pump rate $R_p$ driving population from the ground state $E_0$ into a metastable upper level $E_2$, lasing action from $E_2$ into $E_1$, and ultrafast relaxation from $E_1$ (Fig. 1). Following Siegman [15], the rate equations corresponding to the temporal behaviour of the cavity photon number, $n$, and the $E_2$ population, $N_2$, are as follows:

$$\frac{dn}{dt} = K(n+1)N_2 - \gamma_c n \quad, \tag{1}$$

$$\frac{dN_2}{dt} = R_p - nKN_2 - \gamma_2 N_2 \quad. \tag{2}$$

Here, the population of $E_1$ is assumed approximately null, and $K$ denotes the coupling coefficient for the laser transition, $\gamma_c$ and $\gamma_2$ signifying the cavity and population decay rate, respectively. Under steady-state conditions, equations (1) and (2) may be solved to give the result;

$$n = \frac{R_p - g\gamma_c p + \left[(R_p - g\gamma_c p)^2 + 4\gamma_c R_p\right]^{\frac{1}{2}}}{2\gamma_c} \quad. \tag{3}$$

where $g = \gamma_2/\gamma_{rad}$ and $K = \gamma_{rad}/p$ (in which $\gamma_{rad}$ is the radiative decay rate and $p$ is the number of resonant cavity modes). The relaxation from $E_2$ into $E_1$ is *not* entirely radiative, i.e. $\gamma_2 \neq \gamma_{rad}$, since non-radiative relaxation processes (lattice phonons, wall collisions etc.)

arise. For the present calculational purposes, given that the level $E_2$ decay rate $\gamma_2$ subsumes – but is dominated by – the rate of radiative decay, it is to be assumed that $g = 5/4$ in the absence of the off-resonant input signal considered below. From Eq. (3), employing typical values $p = 10^{10}$ and $\gamma_c = 10^8 \, \text{s}^{-1}$, the familiar ramp in cavity photons at laser threshold emerges, as graphically illustrated by Fig. 2 (solid line).

All-optical control of such a pumped active medium may be achieved by nonlinear optical engagement of the laser emission with stimulated elastic forward scattering of off-resonant (signal) laser pulses, effecting a modification to the dipole transition moment for the $E_2 \to E_1$ laser transition. The mechanism fundamentally entails three matter-photon interactions (Fig. 1); photons annihilated and created into the signal radiation mode (which emerges unchanged) are coupled with the photon emission. The intensity of emission, $I'(\Omega')$, (or power per unit solid angle) follows from Fermi's Rule [16] multiplied by the energy of an emitted photon, $\hbar \omega'$ [17]. Hence, the net intensity is determined from $I'(\Omega') d\Omega' = 2\pi \rho \, \omega' \, |M^{(1)} + M^{(3)}|^2$, where $M^{(1)}$ and $M^{(3)}$ are the quantum amplitudes for the first- and third-order interaction processes, respectively, and $\rho$ is the density of radiation states. The sought optical effects depend on the relative signs of the first- and third-order amplitudes, which are usually of primarily real character; a common sign will lead to emission enhancement, opposite signs its suppression. To proceed, the following detailed result has been determined in recent work [14];

$$I'(\Omega') = \left( \frac{\omega'^4}{8\pi^2 \varepsilon_0 c^3} \right) \Big[ e'_i e'_j \mu_i^{12} \bar{\mu}_j^{12} + (I/c\varepsilon_0) e'_i e_j e_k e'_l \chi_{ijk}^{12}(\omega'; -\omega, \omega) \bar{\mu}_l^{12}$$
$$+ (I^2/4c^2\varepsilon_0^2) e'_i e_j e_k e'_l e_m e_n \chi_{ijk}^{12}(\omega'; -\omega, \omega) \bar{\chi}_{lmn}^{12}(\omega'; -\omega, \omega) \Big] \quad , \qquad (4)$$

where successive terms in square brackets originate from $\left|M^{(1)}\right|^2$, $M^{(1)}\bar{M}^{(3)}$ (plus its conjugate) and $\left|M^{(3)}\right|^2$, respectively. The normal decay transition dipole moment is designated by the shorthand notation $\boldsymbol{\mu}^{12} = \langle 1|\boldsymbol{\mu}|2\rangle$ – in which $|1\rangle$ and $|2\rangle$ denote the states of levels $E_1$ and $E_2$, respectively. Eq. (4) deploys the implied summation convention for repeated (Cartesian) indices, and $I$ is the intensity (irradiance) of the input signal, with $\mathbf{e}'$ and $\mathbf{e}$ representing the linear polarization unit vectors of emission and signal photons, respectively.

The key parameter within Eq. (4) is the nonlinear susceptibility, $\chi_{ijk}^{12}(\omega';-\omega,\omega)$, the explicit form of which is determined from well-attested and reported methods [18,19], and is given by:

$$\chi_{ijk}^{12}(\omega';-\omega,\omega) = \sum_r \sum_{s\neq 2} \left( \frac{\mu_i^{1s}\mu_j^{sr}\mu_k^{r2}}{E_{s2}(E_{r2}-\hbar\omega)} + \frac{\mu_i^{1s}\mu_k^{sr}\mu_j^{r2}}{E_{s2}(E_{r2}+\hbar\omega)} \right)$$
$$+ \sum_r \sum_s \left( \frac{\mu_j^{1s}\mu_i^{sr}\mu_k^{r2}}{(E_{s2}-\hbar\omega+\hbar\omega')(E_{r2}-\hbar\omega)} + \frac{\mu_k^{1s}\mu_i^{sr}\mu_j^{r2}}{(E_{s2}+\hbar\omega+\hbar\omega')(E_{r2}+\hbar\omega)} \right)$$
$$+ \sum_{r\neq 1} \sum_s \left( \frac{\mu_j^{1s}\mu_k^{sr}\mu_i^{r2}}{(E_{s2}-\hbar\omega+\hbar\omega')(E_{r2}+\hbar\omega')} + \frac{\mu_k^{1s}\mu_j^{sr}\mu_i^{r2}}{(E_{s2}+\hbar\omega+\hbar\omega')(E_{r2}+\hbar\omega')} \right),$$

(5)

where $\omega$ is the signal beam frequency, and the transition moments are defined in the same manner as $\boldsymbol{\mu}^{12}$; $r$ and $s$ are intermediate states, equating to either 0, 1 or 2 in the three-level system (except where precluded in certain summations, as indicated above), and $E_{xy} = E_x - E_y$ is an energy difference between two such states. If $\omega' < \omega$, and these frequencies are such that the offset, $\Delta E = E_{20} - \hbar\omega - \hbar\omega'$ is a small fraction of a typical electronic transition, the fourth term of Eq. (5) is dominant so that:

$$e'_i e_j e_k \chi^{12}_{ijk}(\omega'; -\omega, \omega) \approx \frac{\mu^3}{\Delta E(\Delta E + \hbar\omega')} \quad . \tag{6}$$

Here and henceforth, it is assumed that the relevant transition dipole moment components, now simply represented as $\mu$, have broadly similar magnitudes and direction – in calculations on specific systems, this approximation can of course be surrendered for greater accuracy. It should be observed that both factors in the denominator of (6) have negative values, so that the resulting susceptibility is always positive and, thus, denotes enhanced emission; under other conditions the susceptibility components may assume a negative value, representative of reduced emission. On insertion of Eq. (6) into (4), typical values of $I'(\Omega')$ may be calculated for various signal beam intensities.

As indicated above, it is evident that the initial and final terms on the right-hand side of Eq. (4) correspond to spontaneous emission and the nonlinear coupling process, respectively. However, it is the second term (linear in $I$), signifying a quantum interference of these two processes, that represents the leading correction. With this in mind, the degree of enhancement (or in other cases any suppression) of the emission can be measured by taking the ratio of the second term against the first; the corresponding parameter $\eta$ may be approximated as:

$$\eta = \frac{I\mu^2}{c\varepsilon_0 \Delta E(\Delta E + \hbar\omega')} \quad . \tag{7}$$

Returning to Eq. (3), it is clear that the variable *g* will be affected by introduction of the input signal beam, since the radiative decay rate, $\gamma_{rad}$, and population decay rate, $\gamma_2$, both thereby suffer change (but to differing degrees); the non-radiative decay rate, $\gamma_{nr}$, can be assumed to be constant. By simple manipulation, an expression for *g* – dependent on *I* through Eq. (7) – is given by;

$$g = 1 + \frac{1-Y}{Y + \eta Y} \quad , \tag{8}$$

where $Y = \gamma_{rad}/\gamma_2$ and $\gamma_{nr}/\gamma_2 = 1 - Y$. With the previous condition that $g = 5/4$ for $I = 0$, and adopting indicative values $\mu = 16 \times 10^{-30}$ C m, $\Delta E = 10^{-20}$ J and $\hbar\omega' = 10^{-19}$ J, inserting Eq. (8) into (3) generates the results illustrated by two further curves on Fig. 2. Here, the capacity for transistor action is clearly evident. For a constant pumping rate at a level indicated by the dotted vertical line, the system operates below threshold when no signal laser present; however on the introduction of an off-resonant beam with an irradiance approaching $2 \times 10^{11}$ W cm$^{-2}$, the device output climbs by fourteen orders of magnitude, rising to sixteen if the signal input is doubled. Transistor action with respect to the signal beam is clearly exhibited.

The above analysis, based on off-resonant activation of laser emission, represents a new basis for all-optical switching and amplification. The realization of a system suitable for implementing this mechanism is an enticing goal, whose achievement will require the identification of systems for which the key tensor parameters can be optimized; all the necessary theory is now delivered. Our viable all-optical transistor system offers significant advantages over previous schemes including; ultrafast response with high repetition rate, high efficiency, and a straightforward experimental setup. Moreover, it is based on a principle that is not limited to operation with any one specific material; with judicious choice of signal optical frequency, the mechanism is viable in any suitably nonlinear medium.

**Acknowledgement**

The authors thank the Leverhulme Trust for funding this research.

**Figure captions**

Fig. 1.   Energy level diagram of the three-level laser system: black vertical arrows denote electronic transitions, the wavy line represents emission ($\hbar\omega'$), and the off-resonant beam ($\hbar\omega$) is the dashed arrow. The black and open dots symbolize one and two matter-photon interactions, respectively.

Fig. 2.   Plot of log $n$, where $n$ is the number of cavity photons, against the pumping rate, $R_p$, for absent (solid line) and present signal beam; example irradiances of the latter are $2 \times 10^{11}$ W cm$^{-2}$ (dashed) and $4 \times 10^{11}$ W cm$^{-2}$ (dotted). Horizontal arrow illustrates a movement of the lasing threshold to the left for increasing laser intensities. The vertical dotted line represents a constant $R_p$ at which, on introduction of the signal beam, transistor action produces above threshold operation (denoted by the upper pair of horizontal dotted lines).

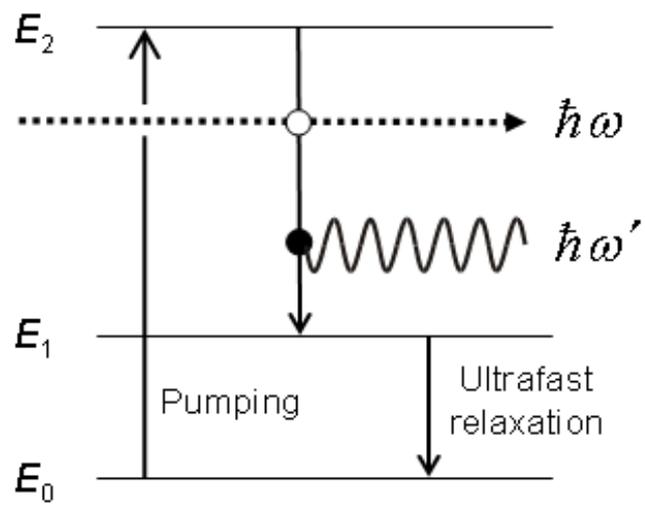

Fig. 1

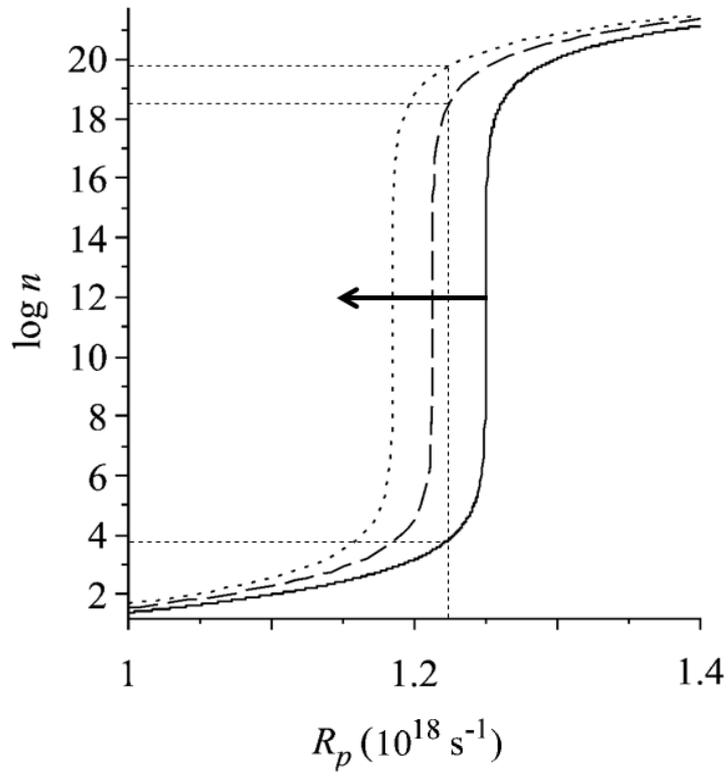

Fig. 2